\documentclass[twocolumn,aps,pra,showpacs,superscriptaddress]{revtex4-1}

\usepackage{amssymb}
\usepackage{mathrsfs}
\usepackage{lipsum}
\usepackage{graphicx}
\usepackage[caption=false]{subfig}
\usepackage{mathtools}
\usepackage{epstopdf}
\usepackage{xcolor}
\usepackage[colorlinks,linkcolor=red,anchorcolor=green,citecolor=blue, urlcolor=violet]{hyperref}
\DeclareGraphicsExtensions{.pdf,.jpg,.png,.eps}
\usepackage[toc,page,title,titletoc,header]{appendix}

\begin{document}

\title{Orbital-angular-momentum-enhanced estimation of sub-Heisenberg-limited angular displacement with two-mode squeezed vacuum and parity detection}

\author{Jiandong Zhang}
\affiliation{Department of Physics, Harbin Institute of Technology, Harbin, 150001, China}
\author{Zijing Zhang}
\email[]{zhangzijing@hit.edu.cn}
\affiliation{Department of Physics, Harbin Institute of Technology, Harbin, 150001, China}
\author{Longzhu Cen}
\affiliation{Department of Physics, Harbin Institute of Technology, Harbin, 150001, China}
\author{Yuan Zhao}
\email[]{zhaoyuan@hit.edu.cn}
\affiliation{Department of Physics, Harbin Institute of Technology, Harbin, 150001, China}

\date{\today}

\begin{abstract}

We report on an orbital-angular-momentum-enhanced scheme for angular displacement estimation based on two-mode squeezed vacuum and parity detection.
The sub-Heisenberg-limited sensitivity for angular displacement estimation is obtained in an ideal situation.
Several realistic factors are also considered, including photon loss, dark counts, response-time delay, and thermal photon noise.
Our results indicate that the effects of the realistic factors on the sensitivity can be offset by raising orbital angular momentum quantum number $\ell$.
This reflects that the robustness and the practicability of the system can be improved via raising $\ell$ without changing mean photon number $N$.

\end{abstract}



\maketitle

\section{Introduction}
Quantum metrology \cite{giovannetti2006quantum, giovannetti2011advances}, as an art and science of precise measurement, is more sensitive than classical metrology, by taking advantage of quantum technologies such as exotic quantum states and detection strategies. 
In the past few decades, the study on the phase estimation in optical interferometers has been a primary subject in quantum metrology \cite{giovannetti2004quantum, escher2011general, barish1999ligo, schnabel2010quantum}. 
Meanwhile, recent progress on quantum measurement theory has played an important role for phase estimation. 
With diverse forms of quantum technologies, the sensitivity of the phase estimation is increased from the shot-noise limit (SNL), the ultimate limit of the classical metrology, to the Heisenberg limit (HL) and even sub-Heisenberg limit \cite{anisimov2010quantum}. 
Quantum metrology has been widely used in different of precision measurement fields, such as gravitational wave detection, optical microscopy and optical lithography \cite{PhysRevLett.104.251102, Israel2017Quantum, boto2000quantum}.

Recently, in addition to phase estimation, considerable amount of researches have been done for angular displacement estimation \cite{zhang2016ultra, liu2017enhancement, PhysRevA.83.053829, PhysRevLett.81.4828, zhang2016super, PhysRevLett.112.200401, cen2017state, d2013photonic, zhang2017improved}. 
Most of these schemes are based on a fact that light can carry two angular momenta: spin angular momentum (SAM) and orbital angular momentum (OAM). 
The Heisenberg-limited sensitivities for both SU(2) and SU(1,1) interferometers have been achieved via squeezed state and entangled state schemes \cite{liu2017enhancement, PhysRevA.83.053829}. 
Generally, these schemes adopt OAM or the combination of OAM and SAM rather than only SAM, which is due to the limitation of the eigenvalue of SAM whereas that of OAM can be an arbitrary integer.

In this paper, we propose an OAM-enhanced estimation scheme using two-mode squeezed vacuum (TMSV) state. 
The angular displacement in our scheme is magnified $2\ell$ times with OAM quantum number $\ell$ in the input state. 
This greatly improves the sensitivity of the estimation, especially when the average photons in the input is limited. 
Additionally, the rotation of the optical axis can be simulated by the angular displacement in our scheme. This suggests that our scheme could be applied to the correction of the reference frames of two coordinated detectors in quantum key distribution (QKD) \cite{d2013photonic}.

The structure of this paper is as follows. In Sec. \ref{II}, we describe our estimation scheme and detection strategy. In Sec. \ref{III}, we compare the resolution and the sensitivity of the ideal case. In Sec. \ref{IV}, we consider several realistic factors in practical estimation process, such as photon loss, dark counts, response-time delay and thermal photon noise. We discuss the enhanced mechanism and give a physical explanation in Sec. \ref{V}. Finally, we summarize our work in Sec. \ref{VI}.

\begin{figure*}[!t]
\centering
\includegraphics[width=\textwidth]{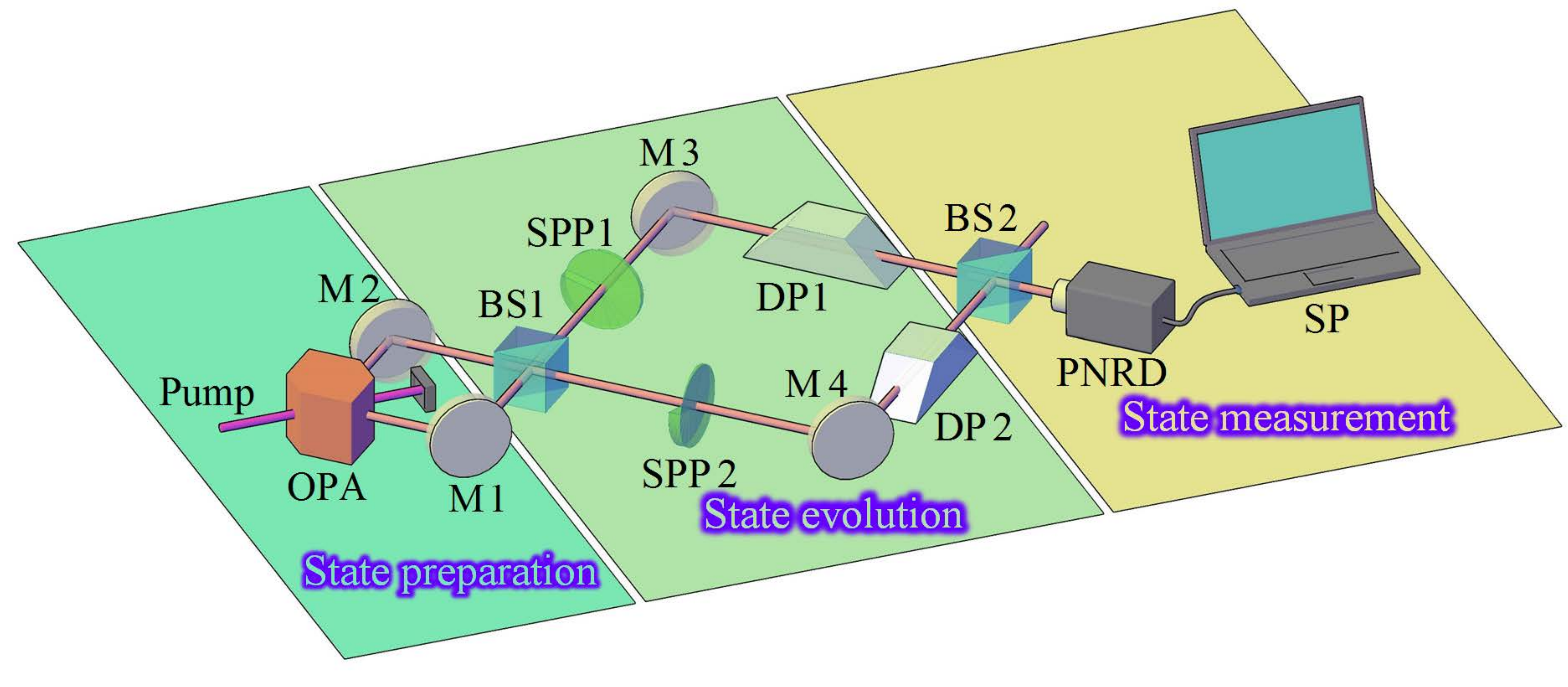}
\caption{Schematic of proposed OAM-enhanced angular displacement estimation. The TMSV state is produced by a OPA and enters the SU(2) interferometer combined with two sets of SPPs and DPs. Parity measurement is carried out in the output. OPA, optical parametric amplifier; M, mirror; BS, beam splitter; SPP, spiral phase plate; DP, Dove prism; PNRD, photon-number-resolving detector; SP, signal processor.}
\label{f1}
\end{figure*}

\section{Enhanced estimation scheme}
\label{II}
Our proposed scheme is shown in Fig. \ref{f1}. The goal of this scheme is to estimate the angular displacement difference between the two Dove prisms (DPs).
The optical parametric amplifier (OPA) plays the role of producing a TMSV state.
Two spiral phase plates (SPPs), two DPs and a SU(2) interferometer constitute the main body of the scheme.
The SPP is used as a modulator where the OAM property could be added to the quantum state. 
The SPPs are placed after the first beam splitter (BS), rather than before it, because the quantum number of OAM turns from $\ell$ to $-\ell$ upon reflection.
The input state enters BS1 then carries OAM via SPPs.
The phase difference $2\ell\varphi $ between the two modes is introduced after the two DPs with angular displacement difference $\varphi$.
Finally, the two modes are coupled by BS2 and parity measurement is carried out in one of the output modes. 

The input state in our scheme is the TMSV state, which has been proved to be the optimal state for estimating phase in a SU(2) interferometer \cite{lang2014optimal}. 
As a vital resource for quantum metrology, quantum computation, quantum communication, and other quantum domains, the TMSV state has also received a great deal of attentions due to its high correlation between two modes \cite{zhang2017effects, anisimov2010quantum}. 
In the Fock state basis, the TMSV state is written as 
\begin{equation}
\left| {{\psi _\textrm{in}}} \right\rangle  = \sum\limits_{m = 0}^\infty  {\sqrt {\left( {1 - t} \right){t^m}} \left| {m,m} \right\rangle },
\end{equation}
where $\left| {m,m} \right\rangle  \equiv {\left| m \right\rangle _{A}} \otimes {\left| m \right\rangle _{B}}$ and $t = {N \mathord{\left/
 {\vphantom {N {\left( {N + 2} \right)}}} \right.
 \kern-\nulldelimiterspace} {\left( {N + 2} \right)}}$. 
Here $N = 2{\sinh ^2}r$ is the mean photon number in the input state, $r$ is the squeezing factor, which is also called the gain strength of OPA.

Parity detection, an excellent detection strategy with binary output, was originally proposed by Bollinger \textit{et al.} \cite{bollinger1996optimal} in the context of trapped ions and adopted in optical metrology by Gerry \textit{et al.} \cite{gerry2000heisenberg, gerry2005quantum}. 
It is a detection strategy that records only the parity of the photon number in the output signal rather than the exact number of photons. 
Take output port $B$ as an example, the parity operator can be written as ${\hat \Pi _{B}} = {\left( { - 1} \right)^{{{\hat b}^\dag }\hat b}}$. 
Consequently, the expectation value of the parity operator has a relationship with the probabilities of even and odd counts, $\langle {{\hat \Pi }_{B}}\rangle  = {P_\textrm{even}} - {P_\textrm{odd}}$. 
In general, parity detection needs photon-number-resolving detector \cite{lundeen2009tomography, cohen2014super}. 
But for Gaussian states, with the aid of the homodyne detection, this process can also be implemented by reconstructing the Wigner function of the output state \cite{plick2010parity}.

\section{Lossless model}  
\label{III}
For Gaussian states, the Wigner function is an effective tool for calculating the output signal. 
It can be constructed by the first moment and second moment of the state given by
\begin{equation}
{W}\left( \textbf{X} \right) = \frac{{\exp \left[ { - {{\left( {\mathbf{X} - \mathbf{M}} \right)}^\top}{\mathbf\Gamma ^{ - 1}}\left( {\mathbf{X} - \mathbf{M}} \right)} \right]}}{{{\pi ^k}\sqrt {\det \left( \mathbf\Gamma  \right)} }},
\end{equation}
where $\mathbf{M}$ is the first moment of the state, also known as the mean, $\mathbf\Gamma $ is the second moment of the state, also known as covariance, $\mathbf{X}$ is vector of phase space variables, and $k$ is the dimension of the state. 
For example, the Wigner function for TMSV state is given by
\begin{widetext}
\begin{equation}
W\left( {{x_1},{p_1},{x_2},{p_2}} \right) = \frac{1}{{{\pi ^2}}}\exp \left[ {2\left( {{p_1}{p_2} - {x_1}{x_2}} \right)\sinh 2r - \left( {x_1^2 + x_2^2 + p_1^2 + p_2^2} \right)\cosh 2r} \right],
\end{equation}
\end{widetext}
where $r$ is the squeezing parameter, and ${x_{1,2}}$ and ${p_{1,2}}$, represents the phase space variables for the two modes. 
Furthermore, we can obtain the variables of the output state by making the replacement
\begin{equation}
{\mathbf{X}_\textrm{out}} = \mathbf{S}{\mathbf{X}_\textrm{in}},
\label{4}
\end{equation}
where $\mathbf{S} = {\mathbf{S}_\textrm{BS2}}{\mathbf{S}_\textrm{AD}}{\mathbf{S}_\textrm{BS1}}$ indicates the propagation process of the SU(2) interferometer, the specific matrix form can be found in Appendix \ref{A}. 
The input column vector, ${\mathbf{M}_\textrm{in}} = {\left( {\begin{array}{*{20}{c}}{{{\left\langle {{\hat x_1}} \right\rangle }_{{\rm{in}}}}} & {{{\left\langle {{\hat p_1}} \right\rangle }_{{\rm{in}}}}} & {{{\left\langle {{\hat x_2}} \right\rangle }_{{\rm{in}}}}} & {{{\left\langle {{\hat p_2}} \right\rangle }_{{\rm{in}}}}}  \\
\end{array}} \right)^\top}$ and the output vector, ${\mathbf{M}_\textrm{out}} = {\left( {\begin{array}{*{20}{c}}{{{\left\langle {{\hat x_1}} \right\rangle }_{{\rm{out}}}}} & {{{\left\langle {{\hat p_1}} \right\rangle }_{{\rm{out}}}}} & {{{\left\langle {{\hat x_2}} \right\rangle }_{{\rm{out}}}}} & {{{\left\langle {{\hat p_2}} \right\rangle }_{{\rm{out}}}}}  \\
\end{array}} \right)^\top}$ represents the phase space variables before and after the propagation through the SU(2) interferometer, respectively.
The expectation values are taken over the input or output state with respect to the subscript in or out. 
The parity signal at the one mode of the output is given by \cite{Gard2017}
\begin{equation}
\label{eq5}
\langle {{{\hat \Pi }}} \rangle  ={\pi }{W_\textrm{out}}\left( {0,0} \right).
\end{equation}
Equation (\ref{eq5}) points out that parity of a single-mode field is equal to the value of its Wigner function at the origin. 


One can show that TMSV state is a Gaussian state with zero mean, ${\mathbf{M}_\textrm{in}} = {( {\begin{array}{*{20}{c}}
   0 & 0 & 0 & 0  \\
\end{array}})^\top}$, and covariance matrix \cite{weedbrook2012gaussian},
\begin{equation}
{\mathbf\Gamma _\textrm{in}} = {\left( {\begin{array}{*{20}{c}}
   {\cosh \left(2r\right) {\kern 1pt} {\mathbf{I}_2}} & {\sinh \left(2r\right) {\kern 1pt}{\mathbf{Z}_2}}  \\
   {\sinh \left(2r\right) {\kern 1pt} {\mathbf{Z}_2}} & {\cosh \left(2r\right)  {\kern 1pt} {\mathbf{I}_2}}  \\
\end{array}} \right)_{4 \times 4}},
\end{equation}
where $ {\mathbf{I}_2}$ is the two-by-two identity matrix and $ {\mathbf{Z}_2}: = {\rm{diag}}\left( {1, - 1} \right)$.
Hence, we can obtain the mean and the covariance of the output state via the following transformations,
\begin{eqnarray}
\nonumber{\mathbf{M}_\textrm{out}} &&= \mathbf{S}{\mathbf{M}_\textrm{in}},\\
{\mathbf\Gamma _\textrm{out}} &&= \mathbf{S}{\mathbf\Gamma _\textrm{in}}{\mathbf{S}^\top}.
\label{7}
\end{eqnarray}
Then we can derive the parity signal of output $B$ which can be written as \cite{ma2017sub}
\begin{eqnarray}
\nonumber\left\langle {{{\hat \Pi }_{B}}} \right\rangle  &&= \frac{{\exp \left( { - \mathbf{M}_{\textrm{out}\left( {3,4} \right)}^\top  \mathbf\Gamma _{\textrm{out}\left( {3,4} \right)}^{ - 1}  {\mathbf{M}_{\textrm{out}\left( {3,4} \right)}}} \right)}}{{\sqrt {\left| {{\mathbf\Gamma _{\textrm{out}\left( {3,4} \right)}}} \right|} }} \\
&&= \frac{1}{{\sqrt {1 + N\left( {N + 2} \right){{\cos }^2}\left( {2\ell\varphi } \right)} }},
\label{8}
\end{eqnarray}
where ${\mathbf{M}_{\textrm{out}\left( {3,4} \right)}} = \left( {\begin{array}{*{20}{c}}
   0  \\
   0  \\
\end{array}} \right)$ and ${\mathbf\Gamma _{\textrm{out}\left( {3,4} \right)}} = \left( {\begin{array}{*{20}{c}}
   {{\gamma _{33}}} & {{\gamma _{34}}}  \\
   {{\gamma _{43}}} & {{\gamma _{44}}}  \\
\end{array}} \right)$.
We can see that the resolution signal of our scheme has a 2$\ell$-fold super-resolution peak from the term ${{\cos }^2}\left( {2\ell\varphi } \right)$ in the denominator of Eq. (\ref{8}).
The matrix elements of the covariance matrix and the specific expression of the parity signal can be found in Appendix \ref{A}. 
On the basis of the expression of the output signal, the visibility \cite{dowling2008quantum, chekhova2016nonlinear}, 
\begin{equation}
V= \frac{{{{\left\langle {{{\hat \Pi }_{B}}} \right\rangle }_{\max }} - {{\left\langle {{{\hat \Pi }_{B}}} \right\rangle }_{\min }}}}{{{{\left\langle {{{\hat \Pi }_{B}}} \right\rangle }_{\max }} + {{\left\langle {{{\hat \Pi }_{B}}} \right\rangle }_{\min }}}}
\end{equation}
can be simplified as $V = {N \mathord{\left/
 {\vphantom {N {\left( {N + 2} \right)}}} \right.
 \kern-\nulldelimiterspace} {\left( {N + 2} \right)}}$. 
Now, we can write the sensitivity using error propagation as 
\begin{equation}\label{sens}
\Delta {\varphi } = \frac{{\sqrt {1 - {{1} \mathord{\left/
 {\vphantom {{} {{R_1}}}} \right.
 \kern-\nulldelimiterspace} {{R_1}}}} }}{{\left| {{{{R_2}} \mathord{\left/
 {\vphantom {{{R_2}} {R_1^{\frac{3}{2}}}}} \right.
 \kern-\nulldelimiterspace} {R_1^{\frac{3}{2}}}}} \right|}},
\end{equation}
where
\begin{eqnarray}
\nonumber{R_1} &&= 1 + N\left( {N + 2} \right){\cos ^2}\left( {2\ell\varphi } \right),\\
{R_2} &&= \ell N\left( {N + 2} \right)\sin \left( {4\ell\varphi } \right).
\label{10}
\end{eqnarray}

The minimum of Eq. (\ref{sens}) could be easily found, when $\varphi  ={\pi \mathord{\left/{\vphantom {1 2}} \right.\kern-\nulldelimiterspace} 4\ell}$, 
\begin{equation}
\Delta {\varphi _\textrm{min}} = \frac{1}{{2\ell\sqrt {N\left( {N + 2} \right)} }}.
\end{equation}
It is a sub-Heisenberg-limited sensitivity, which is boosted by a factor of $2\ell$ with $\ell$ of OAM in the input state. 
So far, this is the best sensitivity for angular displacement estimation in both SU(2) and SU(1,1) interferometers.

\section{Realistic factors}
\label{IV}
The actual detection process is not ideal and some imperfections will exist and affect the estimation results. 
In this section, we discuss the effects of several realistic factors on our model, including photon loss, dark counts, response-time delay and thermal photon noise. 

\subsection{Photon loss}
Photon loss is a realistic factor that cannot be avoided in actual detection process.
This is mainly caused by two processes: photon loss inside the interferometer and photon loss at inefficient detectors.
Both of these processes can be simulated by inserting a fictitious BS in one or two arms of interferometer.
The linear photon loss can be imitated with a relatively simple transformation, on the condition that  all of these states have Gaussian form.
Under these circumstances, the covariance matrix becomes ${\mathbf\Gamma _\textrm{PL}} = \left( {1 - L} \right){\mathbf{I}_4}  {\kern 1pt}\mathbf\Gamma  + L{\kern 1pt}{\mathbf{I}_4}$, where $L$ is the photon loss and ${\mathbf{I}_4}$ is a four-by-four identity matrix. 
Meanwhile, the mean vector can be transformed into ${\mathbf{M}_\textrm{PL}} = \sqrt {\left( {1 - L} \right)} {\kern 1pt}{\mathbf{I}_4}  {\kern 1pt} \mathbf{M}$ \cite{PhysRevA.81.033819}.
Based on the above transformation and Eq. (\ref{7}),
we can derive the expectation value of the output signal as
\begin{equation}\label{13}
{\langle {{{\hat \Pi }_{B}}} \rangle _\textrm{PL}} = \frac{1}{{\sqrt {{K_1}} }},
\end{equation}
and the sensitivity as
\begin{equation}
\Delta {\varphi_{\textrm{PL}}} = \frac{{\sqrt {1 - {1 \mathord{\left/
 {\vphantom {{} {{K_1}}}} \right.
 \kern-\nulldelimiterspace} {{K_1}}}} }}{{\left| {{{{K_2}} \mathord{\left/
 {\vphantom {{{K_2}} {K_1^{\frac{3}{2}}}}} \right.
 \kern-\nulldelimiterspace} {K_1^{\frac{3}{2}}}}} \right|}},
\label{111}
\end{equation}
where 
\begin{eqnarray}
\nonumber{K_1} =&& 1 + \frac{1}{2}{\left( { 1-L } \right)^2}\left\{ {N\left( {N + 2} \right)\cos \left( {4\ell\varphi } \right) + {N^2}} \right\} \\
\nonumber&&+ \left( { 1-{L^2}} \right)N,\\
{K_2} =&& \ell{\left( {1-L} \right)^2}N\left( {N + 2} \right)\sin \left( {4\ell\varphi } \right).
\end{eqnarray}

\begin{figure}[htbp]
\centering
\includegraphics[width=8cm]{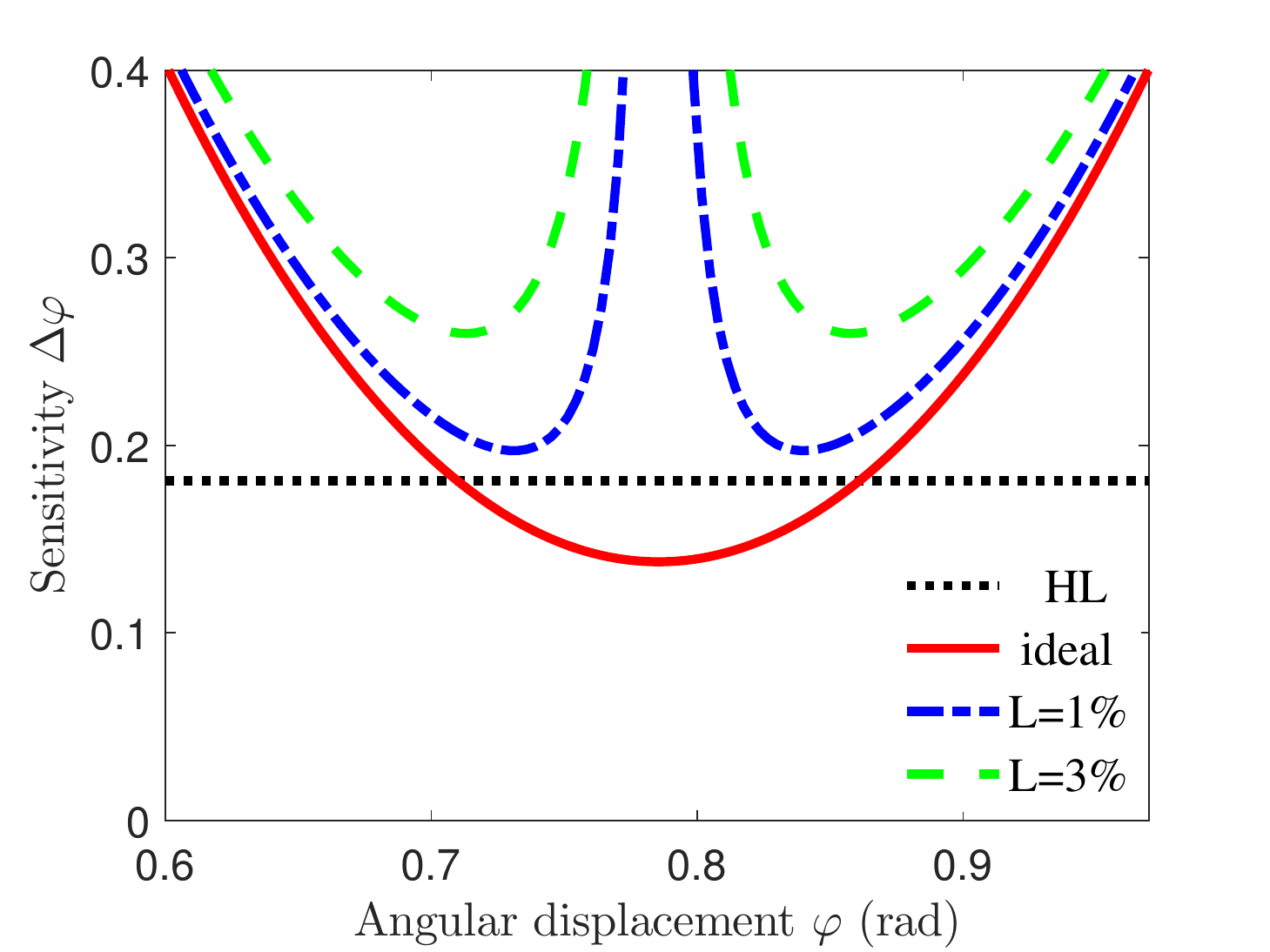}
\centering
\includegraphics[width=8cm]{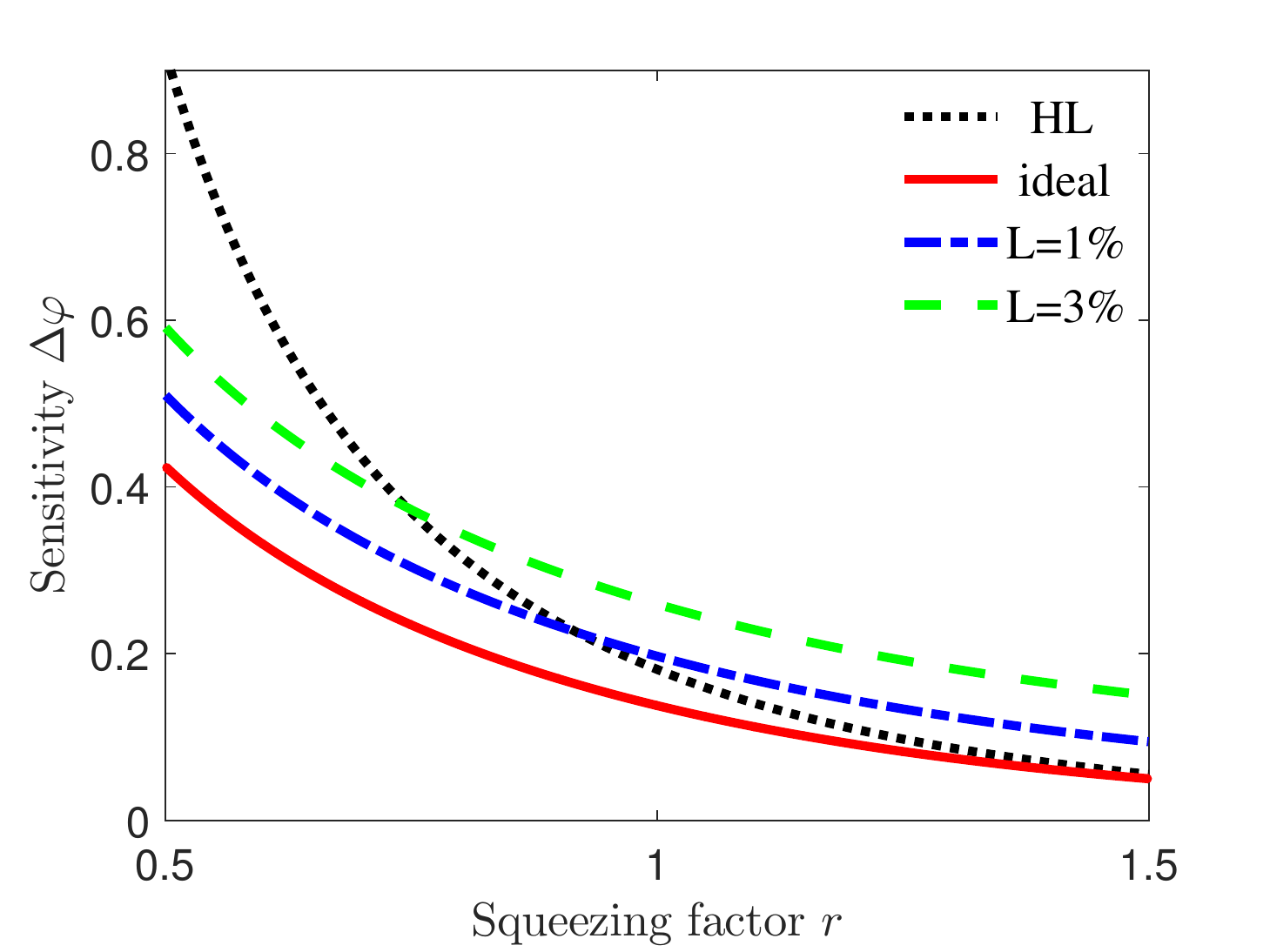}
\caption{(a) Sensitivity with angular displacement and different photon loss values in the case of $\ell=1$ and $r=1$.
(b) Optimal sensitivity of TMSV state with parity detection in the case of $\ell=1$ and photon loss. The losses are $L=1\%$ and $L=3\%$. The value range of squeezing factor $r$ changes from 0.5 to 1.5.}
\label{loss}
\end{figure}

Figure \ref{loss} shows the sensitivity of our scheme with different photon loss values. One can see that the sensitivity is below the HL in the case of low photon loss $L$ and low squeezing factor $r$.
We need to emphasize that in presence of photon loss, it is impossible to achieve sub-Heisenberg limit by increasing the mean photon number $N$. This is due to the fact that the lossless system's sensitivity approximately degenerates into HL for large $N$, where 
\begin{equation}
\mathop {\lim }\limits_{N \to \infty } \frac{1}{{\sqrt {N(N + 2)} }} \simeq \frac{1}{N},
\end{equation}
a slight loss can make the sensitivity worse than the HL. 
Thus, to achieve sub-Heisenberg sensitivity, the system must have low photon loss and small mean photon number.
But from another point of view, the main advantage of our scheme is the $\ell$ in the denominator of Eq. (\ref{13}), which indicates that the sensitivity can be improved by increasing $\ell$ without changing $N$.
So the negative effects of photon loss can be eliminated, i.e., our system is robust to photon loss with the aid of OAM.

\subsection{Dark counts and response-time delay}
Dark counts and response-time delay are the two most common realistic factors that exist in the photon-number-resolving detector. 
Using parameters of a commercial detectors with 5000 APD image elements, where dark counts of each APD less than 100 c/s and the width of sampling gate as 20 ns,
we can deduce that the rate of dark counts to be $d=10^{-2}$.
Response-time delay causes the widening of the sampling gate (usually less than 10 factors) and we choose $d=10^{-1}$ to simulate the existence of both dark counts and response-time delay.

The probability of $w$ dark counts follows the Poisson distribution ${P(w)} = {e^{ - d}}{{{d^w}}}/{{w!}}$.
Thus the output signal with dark counts can be rewritten as \cite{PhysRevA.95.053837}
\begin{equation}
{\left\langle {{{\hat \Pi }_{B}}} \right\rangle _\textrm{DC}} = {e^{ - 2d}}\left\langle {{{\hat \Pi }_{B}}} \right\rangle,
\end{equation}
and the sensitivity is given by
\begin{equation}
\Delta {\varphi _\textrm{DC}} = \frac{{\sqrt {1 - {{{e^{ - 4d}}} \mathord{\left/
 {\vphantom {{{e^{ - 4d}}} {{R_1}}}} \right.
 \kern-\nulldelimiterspace} {{R_1}}}} }}{{{e^{ - 2d}}\left| {{{{R_2}} \mathord{\left/
 {\vphantom {{{R_2}} {R_1^{\frac{3}{2}}}}} \right.
 \kern-\nulldelimiterspace} {R_1^{\frac{3}{2}}}}} \right|}},
\label{222}
\end{equation}
where $R_1$ and $R_2$ have been defined in Eq. (\ref{10}). 

\begin{figure}[!t]
\centering
\includegraphics[width=8cm]{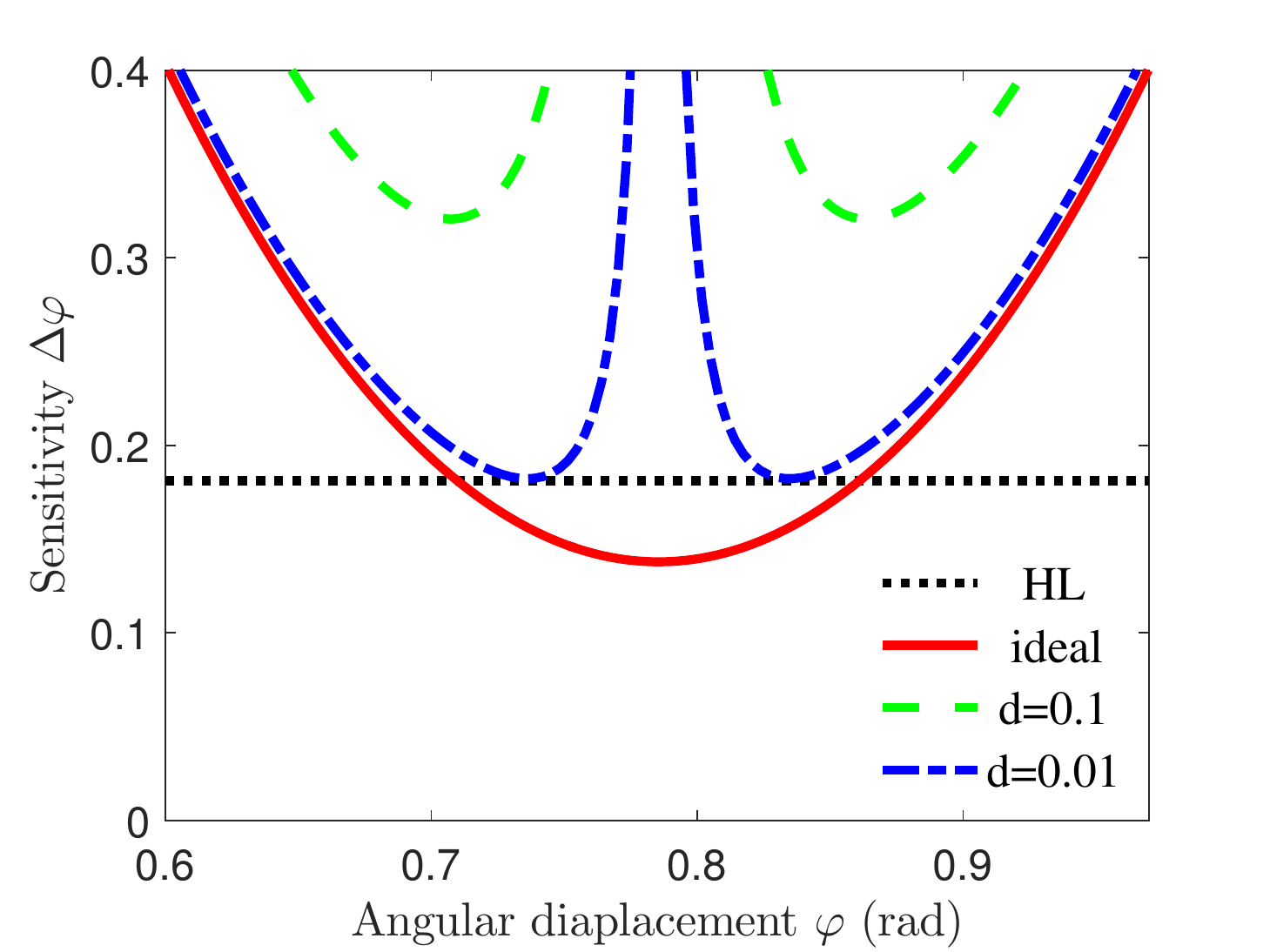}
\centering
\includegraphics[width=8cm]{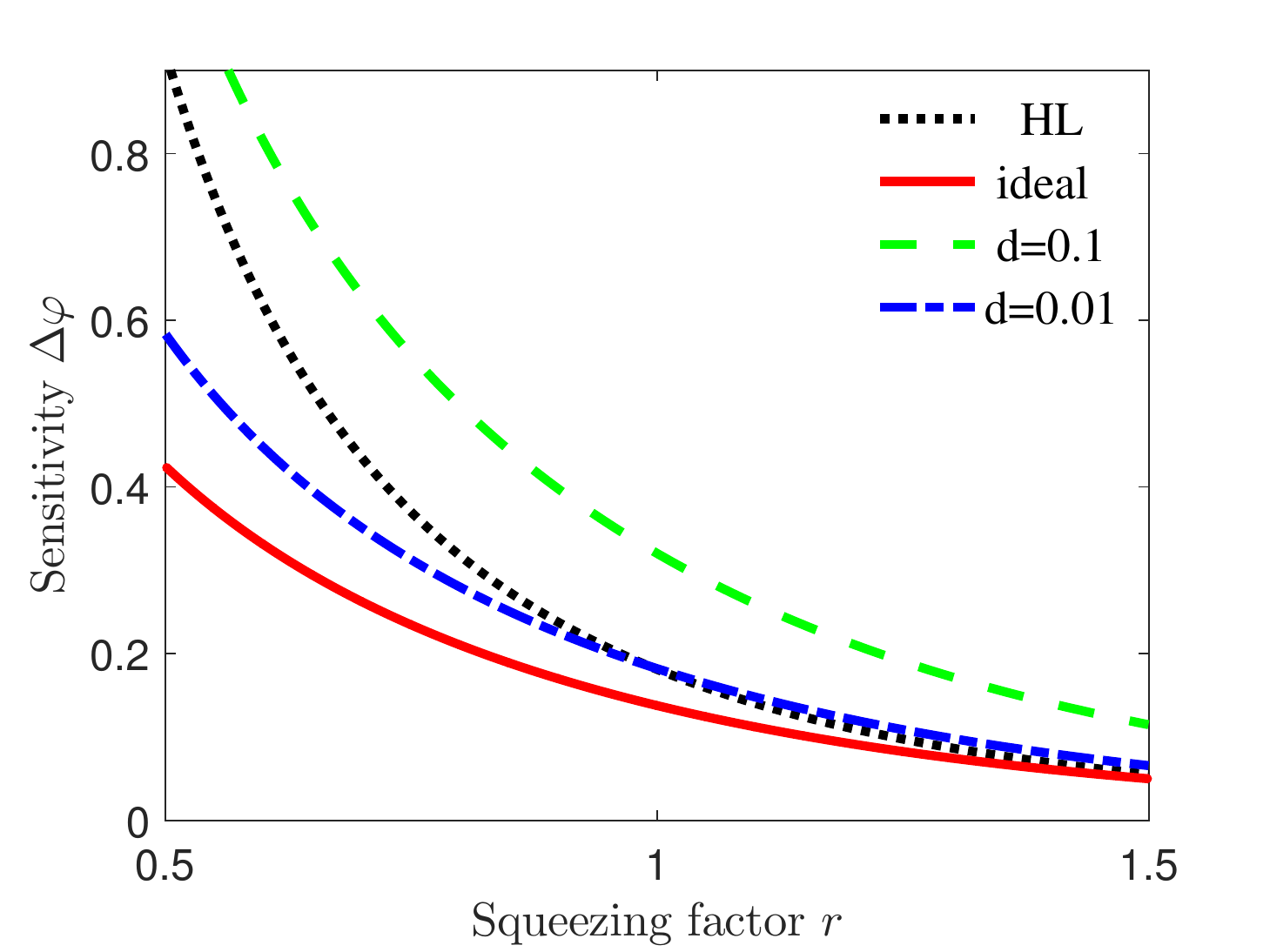}
\caption{(a) Sensitivity with angular displacement, dark counts and response-time delay in the case of  $\ell=1$ and $r=1$. The rate of dark counts $d=0.01$ and $d=0.1$ model the case of only dark counts and response-time delay combined with dark counts, respectively. 
(b) Optimal sensitivity of TMSV state with parity detection in the case of $\ell=1$ and dark counts. The value range of squeezing factor $r$ changes from 0.5 to 1.5.}
\label{dark}
\end{figure}
 
To observe the effects of dark counts and response-time delay on the system's sensitivity, we plot Fig. \ref{dark} with only dark counts and response-time delay combined with dark counts, respectively.
One can see that the effect of dark counts on sensitivity is small, as a result, Heisenberg-limited sensitivity can be obtained even with dark counts.
Further more, response-time delay makes sensitivity worse but not too serious.
This shows that parity detection is robust for dark counts even in the absence of OAM's assistance, thus with the aid of OAM, the stability of the system can be further improved.

\subsection{Thermal photon noise}

Finally, we consider the effect of thermal photon noise, an inevitable interaction with thermal photon from the environment. 
This process is usually implemented by inserting a virtual BS, with system state in one port and thermal state in the other port.
The thermal photon noise at room temperature is approximately $n_\textrm{th}=10^{-20}$, however, $n_\textrm{th}=1$ can be obtained in microwave frequency.
In this section, we place two virtual BSs after two DPs and assume the transmissivities of the virtual BSs are identical.
The thermal state has zero mean vector and its covariance matrix can be found in Appendix \ref{A}.

The output signal can be written as (the detailed calculation can be found in Appendix \ref{B}.)
\begin{equation}
{\left\langle {{{\hat \Pi }_{B}}} \right\rangle _\textrm{TN}} = \frac{1}{{\sqrt {{H_1}} }},
\end{equation}
and using error propagation, the sensitivity is given by
\begin{equation}
\Delta {\varphi _\textrm{TN}} = \frac{{\sqrt {1 - {1 \mathord{\left/
 {\vphantom {1 {{H_1}}}} \right.
 \kern-\nulldelimiterspace} {{H_1}}}} }}{{\left| {{{{H_2}} \mathord{\left/
 {\vphantom {{{H_2}} {H_1^{\frac{3}{2}}}}} \right.
 \kern-\nulldelimiterspace} {H_1^{\frac{3}{2}}}}} \right|}},
\label{333}
\end{equation}
where
\begin{eqnarray}
\nonumber {H_1} =&& \frac{{{T^2}}}{4}\left\{ {2{{\cos }^2}\left( {2\ell\varphi } \right)\left[ {2N\left( {N + 2} \right) + 1} \right] - \cos \left( {4\ell\varphi } \right)}+7 \right\}\\
\nonumber &&+ 1 + 4\left( {n_\textrm{th}^2 + {n_\textrm{th}}} \right){\left( {1 - T} \right)^2} - 2T \\ 
 \nonumber&&+ { 2\left( {2{n_\textrm{th}} + 1} \right)\left( {1 - T} \right)\left( {N + 1} \right)}, \\ 
{H_2} =&& \ell{T^2}\sin \left( {4\ell\varphi } \right) {N\left( {N + 2} \right)}.
\end{eqnarray}
\begin{figure}[htbp]
\centering
\includegraphics[width=8cm]{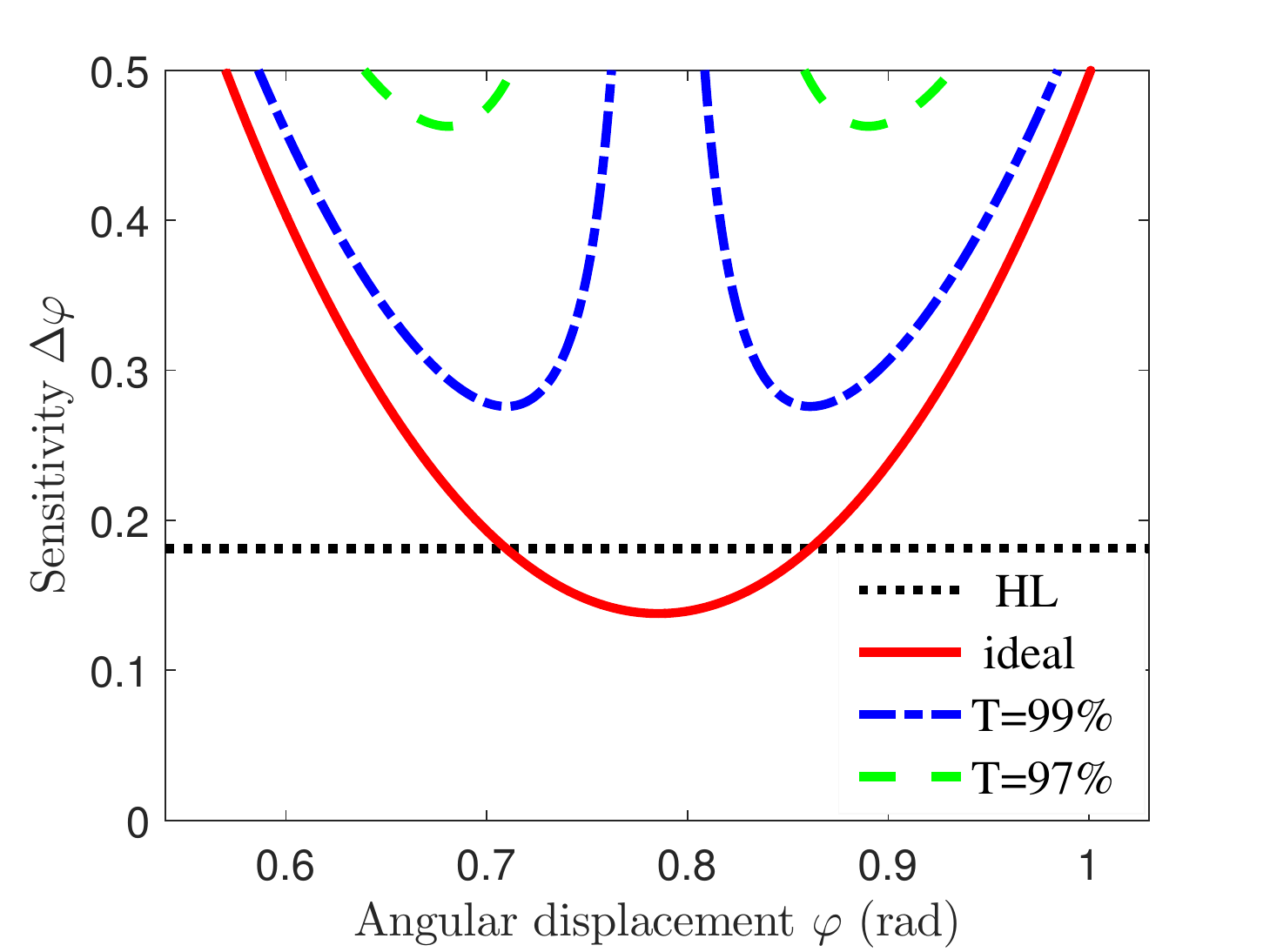}
\centering
\includegraphics[width=8cm]{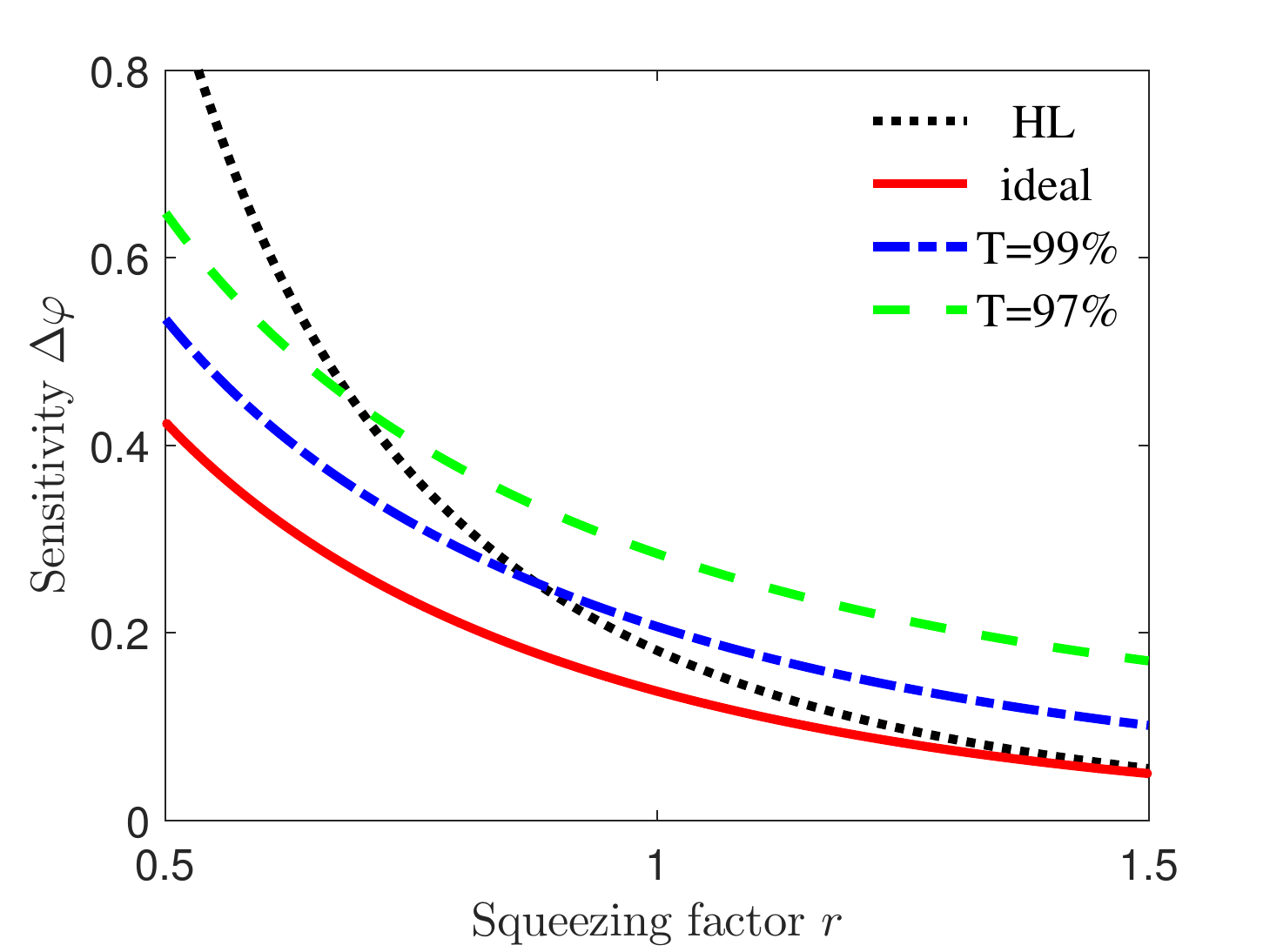}
\caption{(a) Sensitivity with angular displacement and thermal photon noise $n_\textrm{th}=0.1$ in the case of $\ell=1$ and $r=1$. The transmissivities are $T=99\%$ and $T=97\%$.
(b) Optimal sensitivity of TMSV state with parity detection in the case of $\ell=1$ and $n_\textrm{th}=0.1$. The value range of squeezing factor $r$ changes from 0.5 to 1.5.}
\label{noise}
\end{figure}

We plot Fig. \ref{noise} with transmissivities $T=99\%$ and $T=97\%$ for the two virtual BSs. 
One can discover that this situation is similar to that of photon loss, and sub-Heisenberg limit can be realized only by ensuring low squeezing and small transmission loss.
The result is slightly inferior to that of the photon loss case at the same condition, $L=1-T$, for the introduction of thermal noise accompanied by the photon loss of the TMSV state itself.
In addition, a potential advantage of our scheme is that better sensitivity can be achieved by increasing $\ell$ with the same level of thermal photon noise.
This shows that our scheme is of an amplificatory effect for angular sensitivity compared to the schemes without OAM.

\section{Analysis and discussion}
\label{V}
\begin{figure}[htbp]
\centering
\includegraphics[width=8cm]{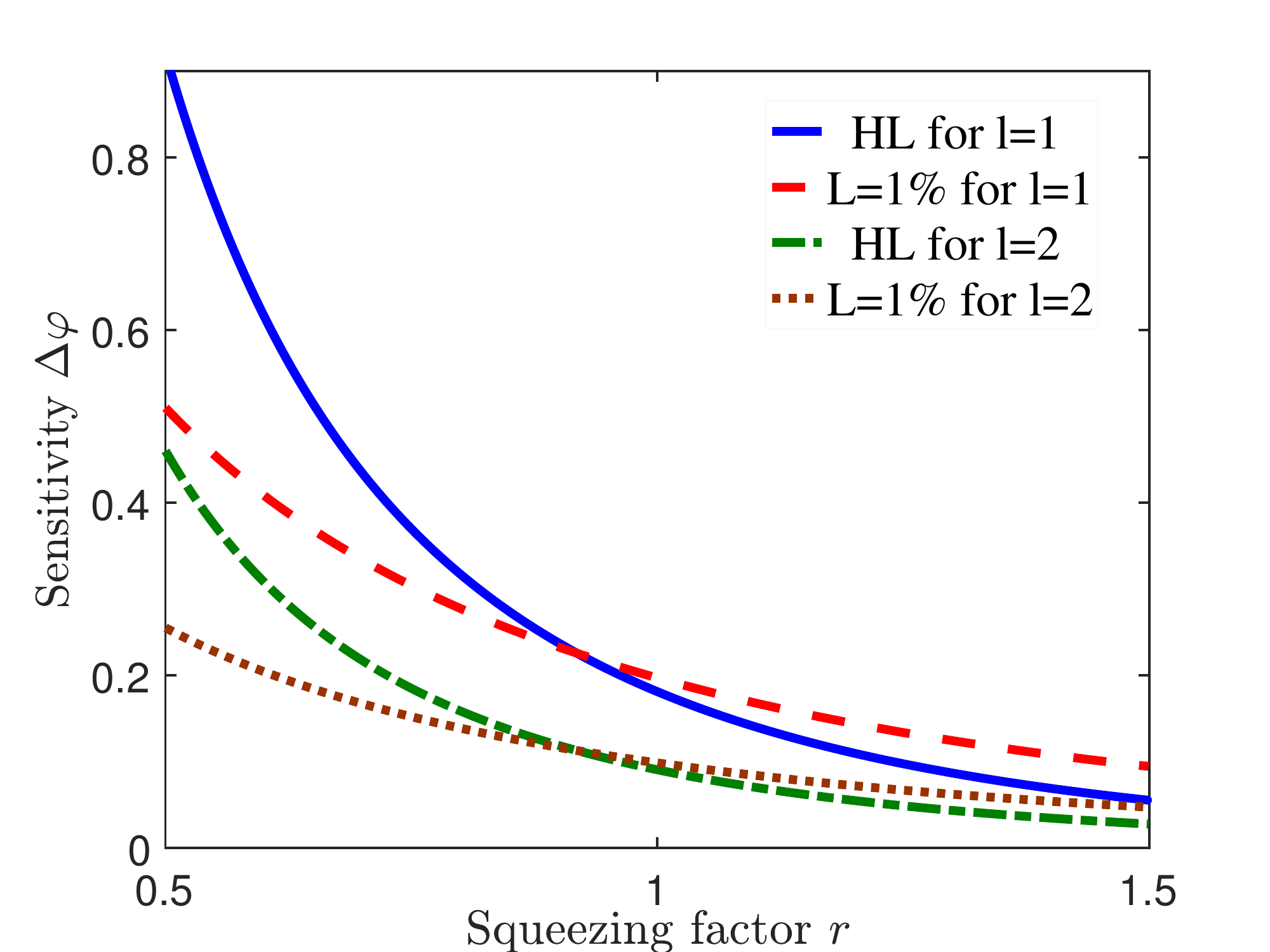}
\caption{Optimal sensitivity of TMSV state with parity detection in the case of  $L=1\%$ and different quantum numbers. }
\label{comparison}
\end{figure}

Here, we discuss the mechanism of enhanced sensitivity with OAM by summing up and analyzing previous results.
For three sensitivities in realistic conditions (Eqs. (\ref{111}), (\ref{222}) and (\ref{333})), the factor $\ell$ can be extracted in the denominator and the sine and cosine terms containing $\ell$ would not change the sensitivity.
Therefore, the increase in sensitivity of OAM quantum number is linear with respect to $\ell$.
In Fig. \ref{comparison}, with the same photon loss $L=1\%$, one can see that the sensitivity of the $\ell=2$ is two times than that of the $\ell=1$.
In general, the sensitivity enhancement of our scheme comes from the linear amplification of $\ell$, when compared to other schemes, with the same photon number and without the use of OAM. 
For a single measurement, in the case of $r=1$ and $L=1\%$, the HL sits at 0.1809, and for our scheme, a sensitivity of 0.1968 can be obtained with $\ell=1$.
The result means the sensitivity is $1.59\times10^{-2}$ higher than HL, whereas the sensitivity is $1.59\times10^{-3}$ better than HL with $\ell=10$ while keeping other parameters the same. This indicates that in the noisy conditions, that with higher $\ell$, our scheme could achieve HL scaling sensitivity. 

From another point of view, raising $\ell$ in our scheme is equivalent to increasing the number of repetitions without changing $N$, i.e., measurement once in our scheme is equivalent to the non-OAM schemes repeated $4\ell^2$ times.
This process is of great significance when the number of measurements to the sample is limited, like biological tissue that can not be exposed for a long time.
Under the current experimental conditions, it is relatively easy to prepare the state with $\ell \le 10$, which suggests that our scheme has an enhanced effect of roughly 1-2 orders of magnitude compared with non-OAM schemes.
Therefore, a linear amplification of a angular displacement or, equivalently, the increase of the statistical effect with the number of repetitions is the enhanced mechanism of our scheme.

\section{conclusion}
\label{VI}
In summary, we propose an OAM-enhanced angular displacement estimation scheme using two-mode squeezed vacuum state and parity detection. 
The result shows a $2\ell$ factor enhancement in sensitivity and a sub-Heisenberg-limited sensitivity under the lossless situation. 
Moreover, a $2\ell$-fold super-resolution is also demonstrated by our scheme. 
As a practical extension, we also study the effects of several realistic factors$\---$photon loss, dark counts, response-time delay and thermal photon noise$\---$on the sensitivity of our scheme.
The results reveal that our scheme is robust to dark counts and response-time delay.  
As to photon loss and thermal photon noise, in the case of low loss, our scheme still provides a considerable sensitivity.
In addition, the effects of realistic factors can be offset by raising $\ell$ without changing other parameters.
Overall, an optimal sensitivity to date for angular displacement estimation is realized, first ever scheme for angular displacement estimation with sub-Heisenberg-limited sensitivity.

\section*{acknowledgments} 
This work is supported by the National Natural Science Foundation of China (Grant No. 61701139).

\appendix

\section{Optical matrices and elements of covariance matrix in an ideal situation}
\label{A}
The transformation matrices of the two BSs and angular displacement are given by 
\begin{equation}
{\mathbf{S}_\textrm{BS1}} = {\mathbf{S}_\textrm{BS2}} = \frac{1}{{\sqrt 2 }}{\left( {\begin{array}{*{20}{c}}
   {{\mathbf{I}_2}} & {{\mathbf{I}_2}}  \\
   {{\mathbf{I}_2}} & {{\kern 1pt}  - {\mathbf{I}_2}}  \\
\end{array}} \right)_{4 \times 4}},
\end{equation}
\begin{equation}
{\mathbf{S}_\textrm{AD}} = {\left( {\begin{array}{*{20}{c}}
   {{\mathbf\Theta _2}} & {{\mathbf{O}_2}}  \\
   {{\mathbf{O}_2}} & {{\mathbf{I}_2}}  \\
\end{array}} \right)_{4 \times 4}},
\end{equation}
where ${\mathbf{O}_2}$ is a two-by-two zero matrix and ${\mathbf\Theta _2}$ is a two-by-two rotation matrix with angular displacement $\varphi$,
\begin{equation}
{\mathbf\Theta _2} = \left( {\begin{array}{*{20}{c}}
   {\cos \left( {2\ell\varphi } \right)} & { - \sin \left( {2\ell\varphi } \right)}  \\
   {\sin \left( {2\ell\varphi } \right)} & {\cos \left( {2\ell\varphi } \right)}  \\
\end{array}} \right).
\end{equation}

The covariance matrix of thermal state is 
\begin{equation}
{\mathbf\Gamma _\textrm{th}} = \left( {2{n_\textrm{th}} + 1} \right) {\mathbf{I}_2}.
\end{equation}

The matrix elements of output covariance matrix are
\begin{equation}
{\gamma _{11}} = {\gamma _{33}} = \cosh \left(2r\right) - {\sin ^2}\left( {2\ell\varphi } \right)\sinh \left(2r\right),
\end{equation}
\begin{equation}
{\gamma _{22}} = {\gamma _{44}} = \cosh \left(2r\right) + {\sin ^2}\left( {2\ell\varphi } \right)\sinh \left(2r\right),
\end{equation}
\begin{eqnarray}
\nonumber{\gamma _{12}} =&& {\gamma _{21}} = {\gamma _{14}} = {\gamma _{41}} = {\gamma _{23}} = {\gamma _{32}} = {\gamma _{34}} = {\gamma _{43}} \\
=&& \frac{1}{2}\sin \left( {4\ell\varphi } \right)\sinh \left(2r\right),
\end{eqnarray}
\begin{equation}
{\gamma _{13}} = {\gamma _{31}} = {\cos ^2}\left( {2\ell\varphi } \right)\sinh \left(2r\right),
\end{equation}
\begin{equation}
{\gamma _{24}} = {\gamma _{42}} =  - {\cos ^2}\left( {2\ell\varphi } \right)\sinh \left(2r\right).
\end{equation}

\section{Optical matrices, elements of mean and covariance matrix in thermal state coupling}
\label{B}
The calculation of thermal state coupling is similar to that of ideal situation.
For the conservation of photon number, we need to consider the environment modes.
Hence, the system mode  turns from two modes to four modes and  the previous four-by-four matrices are replaced by eight-by-eight matrices.
The matrices of input, BSs and angular displacement are given by 
\begin{equation}
\mathbf\Gamma _\textrm{in}^ * = {\left( {\begin{array}{*{20}{c}}
   {{\mathbf\Gamma _\textrm{in}}} & {{\mathbf{O}_4}}  \\
   {{\mathbf{O}_4}} & {{\mathbf\Gamma _\textrm{th}} \oplus {\mathbf\Gamma _\textrm{th}}}, \\
\end{array}} \right)_{8 \times 8}},
\end{equation}
\begin{equation}
\mathbf{S}_\textrm{BS1}^ *  = \mathbf{S}_\textrm{BS2}^ *  = {\left( {\begin{array}{*{20}{c}}
   {{\mathbf{S}_\textrm{BS}}} & {{\mathbf{O}_4}}  \\
   {{\mathbf{O}_4}}& {{\mathbf{I}_4}}  \\
\end{array}} \right)_{8 \times 8}},
\end{equation}
\begin{equation}
\mathbf{S}_\textrm{AD}^ *  = {\left( {\begin{array}{*{20}{c}}
   {{\mathbf{S}_\textrm{AD}}} & {{\mathbf{O}_4}}  \\
   {{\mathbf{O}_4}} & {{\mathbf{I}_4}}  \\
\end{array}} \right)_{8 \times 8}},
\end{equation}
where ${\mathbf{O}_4}$ is a four-by-four zero matrix. 
The matrix of the virtual BS is
\begin{equation}
\mathbf{S}_\textrm{VBS}^ * = {\left( {\begin{array}{*{20}{c}}
   {\sqrt T  {\kern 1pt}{\kern 1pt}{\mathbf{I}_4}} & {\sqrt {1 - T}   {\kern 1pt}{\kern 1pt}{\mathbf{I}_4}}  \\
   {\sqrt {1 - T}  {\kern 1pt}{\kern 1pt} {\mathbf{I}_4}} & { - \sqrt T   {\kern 1pt}{\kern 1pt}{\mathbf{I}_4}}  \\
\end{array}} \right)_{8 \times 8}},
\end{equation}
where $T$ is the transmissivity of the virtual BS.

The whole transformation relation can be written as 
\begin{equation}
\mathbf\Gamma _\textrm{out}^ *  =  {\mathbf{S}^ *}\mathbf\Gamma _\textrm{in}^ * {\left( {\mathbf{S}^ * } \right)^\top},
\end{equation}
where ${\mathbf{S}^ * }={\mathbf{S}_\textrm{BS2}^ * \mathbf{S}_\textrm{VBS}^ * \mathbf{S}_\textrm{AD}^ * \mathbf{S}_\textrm{BS1}^ * } $.

Similarly, the dimension of the mean vector also changes from one-by-four to one-by-eight.
Since the thermal state has zero mean vector, the mean vector becomes
\begin{equation}
\mathbf{M}_{{\rm{in}}}^ *  = {\left( {\begin{array}{*{20}{c}}
   {{\mathbf{M}_{{\rm{in}}}}} & {{\mathbf{M}_{{\rm{th}}}}}  \\ 

\end{array}} \right)^\top}.
\end{equation}
Therefore, both the input and the output mean vectors become one-by-eight zero vectors.


%

\end{document}